\documentclass[12pt,a4paper]{article}
\usepackage{graphicx}
\newcounter{N}
\usepackage{verbatim}
\begin{document}

\begin{center}
{\bfseries Tubular initial conditions and ridge formation}
\vskip 5mm
M.S. Borysova$^{\ddag}$, O.D. Borysov$^{\flat}$, Iu.A. Karpenko$^{\dag,2}$, V.M. Shapoval$^\dag$ and Yu.M. Sinyukov$^\dag$,
\vskip 5mm
{\small {\it $^\dag$ Bogolyubov Institute for Theoretical Physics, Kiev, 03680, Ukraine}}\\
{\small {\it $^\ddag$ Kyiv Institute for Nuclear Research, Kiev,  03083, Ukraine}}\\
{\small {\it $^2$ FIAS, Ruth-Moufang-Str.~1, 60438 Frankfurt am Main, Germany}}\\
{\small {\it $^\flat$ FNAL, Batavia, IL,60510, USA}}\\
\end{center}
\vskip 5mm

\centerline{\bf Abstract}
The 2D azimuth \& rapidity structure of the two-particle  correlations in relativistic A+A collisions is altered significantly by the presence of sharp inhomogeneities in superdense matter formed in such processes. The causality constraints enforce one to associate the long-range longitudinal correlations observed in a narrow angular interval, the so-called (soft) ridge, with  peculiarities of the initial conditions of collision process. This study's objective is to analyze whether multiform initial tubular structures, undergoing the subsequent hydrodynamic evolution and gradual decoupling, can form the soft ridges. Motivated by the flux-tube scenarios, the initial energy density distribution contains the different numbers of high density tube-like boost-invariant inclusions that form a bumpy structure in the transverse plane.   The influence of various structures of such initial conditions in the most central A+A events on the collective evolution of matter, resulting spectra, angular particle correlations and $v_n$-coefficients is studied in the framework of the HydroKinetic Model (HKM). 

PACS number(s): 25.75.Gz, 24.10.Nz
\vskip 10mm

\section{Introduction}

The correlation and fluctuation study is a valuable tool for probing the dynamics of heavy-ion collisions, the initial conditions, and for exploring the thermodynamic properties of strongly interacting matter at extremely high temperatures and/or densities. The data from the experiments at RHIC revealed interesting features in the   two-particle correlation landscape \cite{Horner, Alver, Alver2, Abelev, Adare, Wosiek}. Specifically, an excess of correlated particles in a wide pseudorapidity interval and a narrow azimuthal angle range $\Delta\varphi$ was firstly measured in the relativistic heavy-ion collisions by the STAR collaboration \cite{Horner}. The pair correlation on the near side of the trigger particle (with $p_T= 2.5-15$ GeV) was found to be extended across the entire detector pseudorapidity acceptance region $\Delta\eta\sim 3-4$ units. Such a correlation structure was called ``the ridge''. The ridges are observed both in the correlations of particles with a jet triggering (the ``hard'' ridge) and in the correlations without a high $p_T$ trigger particle (the ``soft'' ridge). The analyses of measurements by the PHENIX \cite{Adare} and PHOBOS \cite{Wosiek} RHIC collaborations confirmed the STAR results. In ALICE LHC experiment it was found that the soft ridge is consistent with expectations from collective response to anisotropic initial conditions \cite{ALICE}. Moreover, the recent measurements by CMS collaboration \cite{CMS1,CMS2} detected unexpected similar effect in  proton-proton collisions at LHC. The clear and significant ridge structure emerges at $\Delta\varphi\approx 0$, extending to $\left|\Delta\eta\right|$ of at least 4 units. This novel feature of the data has never been seen in two-particle correlation functions in $pp$ or $p\bar p$ collisions before.

The discovery of the ridge structures has aggravated quantitative theoretical analyses of nucleus-nucleus and proton-proton collisions but brought the new physical ideas. Early models of the ridge formation were based on the opposite physical mechanisms. Some authors treated the ridge as an initial-state effect \cite{Dumitru}; the others explored final-state effects, soft interactions and jet propagation in anisotropic plasma, as the origin of the ridge \cite{Schenke}. In the former case the authors argued that due to causality constraints the long-range correlations within 4 units of pseudorapidity could be explained only if they originated from the very initial collision stage.  They could be a consequence of the correlations in the classical color fields responsible for multi-particle production in relativistic heavy-ion collisions. Then, due to fluctuations of color charges in colliding nuclei, the longitudinally boost-invariant and transversally bumpy structure of the matter can be formed.  As for the hard ridges, these correlations are associated with jet propagation, that gives relatively narrow correlations in pseudorapidity. The interesting attempt to combine soft and hard ridge structures is done in \cite{Werner}. It is based on interaction of jets with pieces of expanding bulk matter boosted by the transverse flow.

The role of fluctuations in formation of the soft observables was first considered in \cite{Hama0,Hama}, and their ability to produce ridges in \cite{Hama2}.  Alver and Roland \cite{Roland} observed that the correlations arising from geometrical fluctuations in the initial conditions result in odd flow harmonics such as a triangular flow $v_{3}$.  Then one can suppose that $v_{3}$, together with flow harmonics of all higher orders could explain the excess in the long-ranged two-particle correlations. It has been recently  shown in measurements \cite{CMS1,CMS2,Lacey,Grosse,Jia} and demonstrated in various modelings \cite{Ollitrault,Luzum,Bass} that there are fairly strong fluctuations in the initial matter profile from event to event. Such fluctuations contain various higher order harmonics in azimuthal angle $\sim cos (n\phi)$. After the collective expansion of the bulk matter these fluctuations lead to observed harmonic flows up to about $n = 6$. There is a hope that such high harmonics can explain the soft di-hadron azimuthal correlations -- the soft ridge. It means, in fact, that the soft ridges originate from the fluctuating initial conditions altogether with the specific set of Fourier $n$-harmonics.
It also seems, that the fluctuations of the initial conditions, associated with ridges, are of specific type. This issue is studied in the present article.

If the soft ridge  is caused by the peculiarities of the initial space structure of the bulk matter, then this effect should be analyzed within hydrodynamics models of A+A collisions, which are now the standard approach to the description of such processes. During more than 50 years the hydrodynamic models were based on the smooth initial energy density profiles (see, e.g. \cite{Gavin,Hirano} for the ridge problem). However, it has been shown, that fluctuations in the positions of nucleons within the colliding nuclei from event to event may lead to significant deviations from the smooth regular profiles in one event \cite{Manly}. The similar effect can be caused by the fluctuations of the local color charge in Color Glass Condensate effective field theory \cite{Gavin1}. This irregular structure of the initial conditions for hydrodynamic evolution was explored in \cite{Hama2, Petersen, Karpenko, Andrade} for analysis of the ridge phenomenon. It becomes clear, that the most important factor for ridge formation is not just the variation of the geometry form of initial system from event to event, but strongly inhomogeneous bumpy structure of the initial energy density profile in the transverse plane. Such initial structure is subjected to further non-trivial evolution during the system expansion. The different mechanisms can be responsible for the formation of the bumps in the initial energy density profile, e.g. it can be longitudinal strings as in Nexus \cite{Hama2}, color flux tubes, arising between large locally fluctuating color charges in colliding nuclei. These fluctuations have the form of very narrow and dense, approximately  boost-invariant longitudinal tubes, shifted differently from the center in transverse plane, on the top of smoothly distributed energy density.  The number of the tubes also can be different. The  analysis of the ridge formation in the case of one peripheral tube has been done in \cite{Hama2, Andrade} with the Cooper-Frye prescription for freeze-out. The analogous analysis for multi-tube systems was provided in \cite{Borysova, multi}.    

In this work we continue to analyze possibilities of ridge formation at sharp disturbances of the initial energy density. The evolution and spectra formation for the system with bumpy initial energy density profile is analyzed within the HydroKinetic Model (HKM)\cite{Sinyukov, Akkelin, Sinyukov2, Sinyukov3} which incorporates description of all the stages of the system evolution, including the afterburn one, responsible for continuous particle liberation from expanding decoupling system. Previously it was found that the effect of the initial bump-like fluctuations is not washed out during the system expansion and leads to the specific final energy density distributions \cite{Borysova,Borysova2,Borysova3}. In this paper we examine the influence of tube-like fluctuations of different type on the observed particle spectra, azimuthal correlations and magnitudes of the Fourier harmonics. To highlight the role of the bumpy structure, we limit our present study to the most central collisions only, say $c=0-2\%$.  In contrast to the non-central collisions, with strong initial eccentricity already in average (in the framework of the variable geometry analysis) and large $2^{nd}$ flow harmonic produced, in the perfectly central collisions $(b = 0)$ the  background geometry is isotropic, and the anisotropy and $v_n$-coefficients, caused by the fluctuating bumpy structure only, will be best manifested.

\section{Model description}

The hydrokinetic approach, which we use as the basic model of the matter evolution, is described in detail in \cite{Akkelin,Sinyukov2}. It contains the perfect hydrodynamic component, related to expanding quark-gluon matter, and the kinetic one, describing the system decay and the spectra formation due to gradual particle liberation from expanding hadron-resonance matter.  We consider the transversally bumpy, tube-like IC aiming to study how the initial fluctuations in the energy density distribution affect the spectra, azimuthal correlations and flows.  For the sake of simplicity, these results are presented  only for one kind of particles -- negative pions $\pi^-$.  

The HKM in its original version  \cite{Akkelin,Sinyukov2} is (2+1)D model based on the boost-invariant Bjorken-type IC, where the longitudinal flow has quasi-inertial form and is related to the initial proper time $\tau_i=\sqrt{t^2-z^2}$, when (partial) thermalization is established and further evolution can be described by hydrodynamics. The transverse dynamics of the prethermal stage starts at very early time $\tau_0 =0.1-0.2$~fm/c just after the c.m.s. energy in the overlapping region of colliding nuclei exceeds their binding energy. The thermalization can hardly happen before 1 fm/c (1--1.5 fm/c is the lowest known estimate of the thermalization time). However, if one starts hydrodynamics at that time, neither radial nor elliptic flow can develop well enough to explain experimental transverse spectra and their anisotropy ($v_2$-coefficients). The solution of this problem was proposed in the papers \cite{Sinyukov3,Sinyukov4}, where it was demonstrated how efficiently the transverse collective flow and its anisotropy can develop at prethermal stage (even without pressure) in spatially finite systems typical for A+A collisions. Now the description of this pre-thermal stage, based on the Ref. \cite{Akkelin2} is in progress. Meanwhile, in \cite{KarpSinWern} a rough approach was proposed, which allows one to calculate the developing of the transverse flow and its anisotropy at the pre-thermal stage by means of hydro evolution that starts at very early time (0.1 fm/c)  just to account for the energy-momentum conservation law. It brings a good agreement with all bulk observables. It does not mean that thermalisation actually happens at the proper time 0.1--0.2~fm/c \footnote{the same concerns the hydrodynamic models where the starting time is 0.4--0.6 fm/c}, but only that hydrodynamic approach introduces no big errors being applied out of its applicability region, at the prethermal stage, to describe collective flow and its anisotropy at the thermalization time $\sim 1$~fm/c. With this in mind the hydrodynamic evolution in our approach begins from the starting time of the collision process, $\tau_i=\tau_0$. Here we use the starting proper time $\tau_0 = 0.2$~fm/c. 
At this very early moment there is no collective transverse flow, and our analysis is related just to this case.

We assume the initial energy density distribution in the corresponding flux-tubes (produced by the fluctuating local nucleons distribution or color charge fluctuations in colliding nuclei) to be fairly homogeneous in a long rapidity interval because of the boost-invariance and rather thin transversally with the transverse (Gaussian) radii $a_i = 1$ fm. The initial energy-density distribution $\epsilon(x,y) $ at $\tau_0$ is supposed to be decomposed in general case as

\begin{eqnarray}\label{edd}
\epsilon(x,y) = \epsilon_{bkg}(r)+\sum\limits_{i=0}^{N_t}  \epsilon_i e^{-\frac{(x-x_i)^2+(y-y_i)^2}{a^2}},
\end{eqnarray}
where $r^2=x^2+y^2$, ${\bf r}_i=(x_i,y_{i})$  are the positions of the tubes' centers, $N_t$ is the number of tubes, $\epsilon_i$ are the values of maximal energy density in the tube-like fluctuations.
In our  analysis we use the two different types of the background $\epsilon_{bkg}$, on which tubes are placed: it takes either the Gaussian form 
\begin{eqnarray}\label{edd1}
\epsilon_{bkg}^G(r)=\epsilon_{b}e^{-\frac{r^2}{R^2}},
\end{eqnarray}
or the Woods-Saxon form with the surface thickness parameter $\delta$
\begin{eqnarray}\label{edd2}
\epsilon_{bkg}^{WS}(r)=\frac{\epsilon_{b}}{e^{\frac{\sqrt{r^2} -R_a}{\delta}}+1},
\end{eqnarray}
 where $\epsilon_b$ is the maximum value of the background energy-density distribution.

The results are demonstrated for various kinds of the bumpy IC structure with different number $N_t$ of tubes -- two odd ones ($N_t$ = 1 and 3) and two even ones ($N_t$ = 4 and 10) having variable radial distances from the center. Also the two previously mentioned types of the background are utilized. Specifically, the analysis is carried out for the following tube-like initial configurations:

\begin{list}{\roman{N}}{\usecounter{N}} 
\item The configuration with a smooth Gaussian profile without fluctuations, as it was considered in \cite{Sinyukov3}, where the parameters obtained from the fit to the Color Glass Condensate model result are as follows: $R = 5.4$ fm and maximum energy density at $r = 0$ is $\epsilon_{b} = 90 $ GeV/fm$^3$. 
\item The configuration with one tube  in the center, where the tube energy density profile is the Gaussian one with $a = 1.0$~fm and $\epsilon_i=270$~GeV/fm$^3$. The background corresponds to the case (i).
\item  The configuration with one tube shifted from the center, where $\epsilon_i = 270$~GeV/fm$^3$, $r_1 = 3$~fm or $r_1 = 5.6$~fm, and $a = 1$~fm. The background corresponds to item (i). The initial configuration with $r_1 = 3$~fm is presented in Fig.~1 (left).
\item The configuration with three tubes: $\epsilon_i= 250$~GeV/fm$^3$; $\textbf{r}_i = (0, 5.6)$, $(-1, 3.6)$, $(-1, -3.6)$~fm or $\textbf{r}_i =(0, 0)$, $(-1, 3.6)$, $(-1, -3.6)$~fm; $a_i = 1$~fm. The background is Gaussian with $\epsilon_{b} = 85$~GeV/fm$^3$, $R = 5.4$~fm.
\item The configuration with four symmetrically located tubes: $\epsilon_i=250$ GeV/fm$^3$, $r_i = 5.6$~fm and $a_i = 1$~fm. The background is the Gaussian one with $\epsilon_b=85$~GeV/fm$^3$. The corresponding initial energy density profile is presented in Fig.~1 (center).
\item The configuration with ten tubes, $\epsilon_{b} = 25$~GeV/fm$^3$, $R = 5.4$~fm, $r_1 = 0$~fm, $r_{2,3,4} = 2.8$~fm, $r_{i> 4} = 4.7$~fm, $a = 1$~fm, and $\epsilon_i=4\epsilon_{b}$ (see  Fig.~1 right). 
\item Besides these configurations we consider also the case of IC with background profile $\epsilon_{bkg}(r)$ in the form (3) ($\epsilon_{b} = 90$~GeV/fm$^3$, $R = 6.37$~fm, and $\delta = 0.54$~fm) and one tube shifted from the center and placed at $r_i = 0.5R$ or $1.1R$. The maximum energy densities are $\epsilon_i = 270$ GeV/fm$^3$, $a = 1$~fm. This type of IC with $r_i = 0.5R$ is presented in Fig.~2.
\end{list}

\begin{figure}[!htb]
\minipage{0.32\textwidth}
  \includegraphics[width=\linewidth]{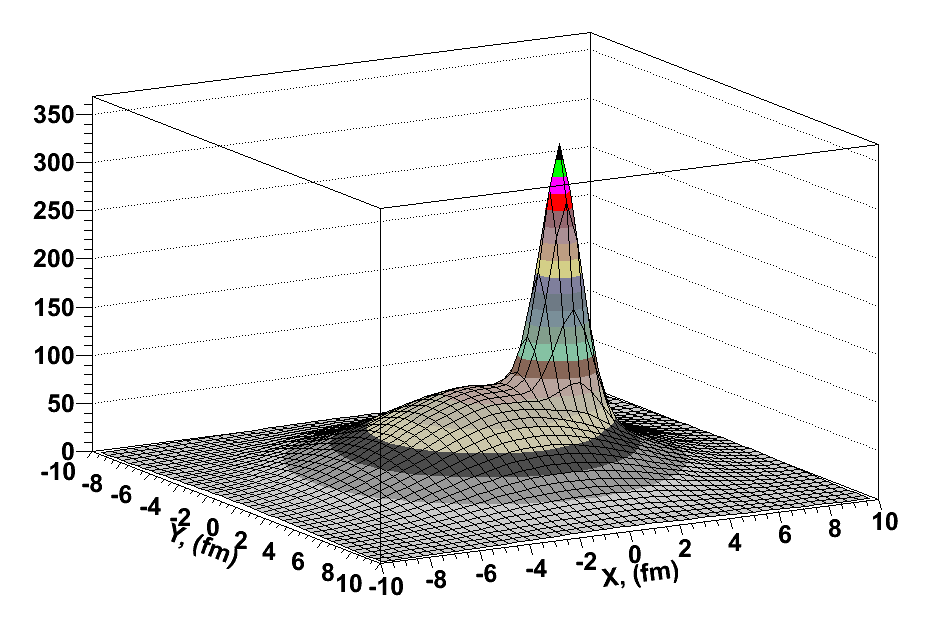}
 \endminipage\hfill
\minipage{0.32\textwidth}
  \includegraphics[width=\linewidth]{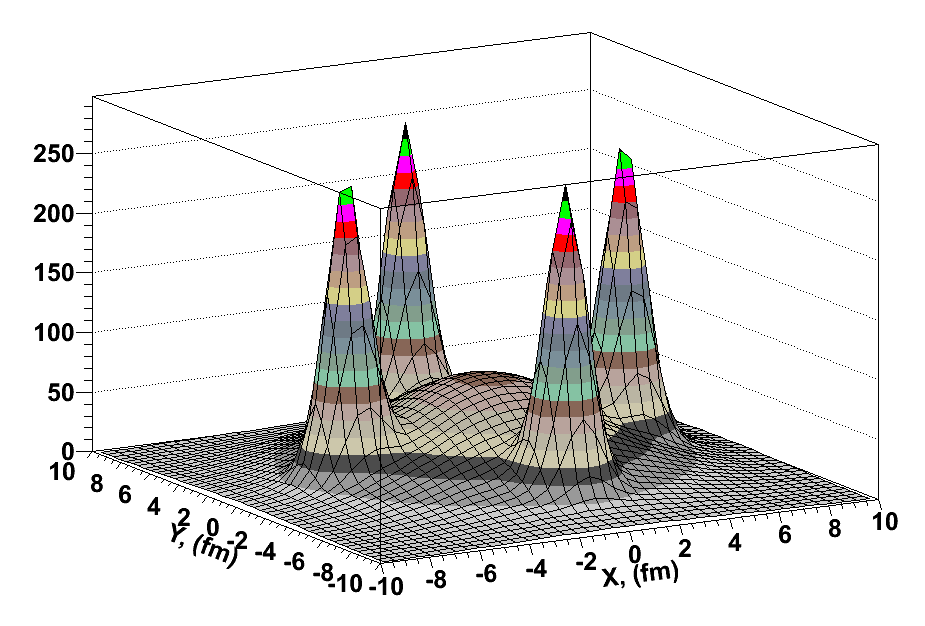}
  \endminipage\hfill
\minipage{0.32\textwidth}%
  \includegraphics[width=\linewidth]{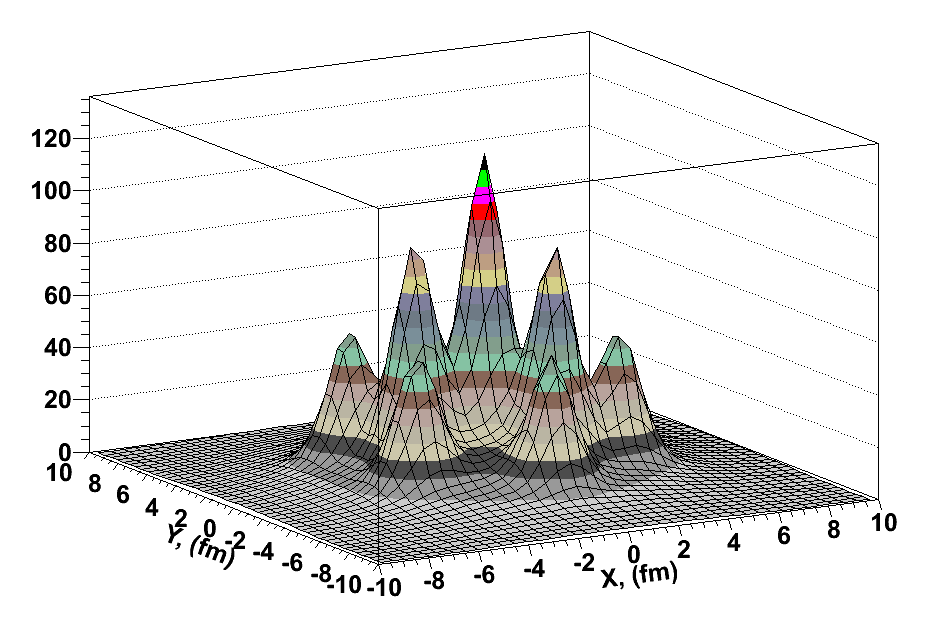}
  \endminipage
\caption{3D plots of the initial energy density profiles with tube-like IC on the background in the form (\ref{edd1}) for $\tau_0 = 0.2$~fm/c.  Left -- 1 tube (iii) displaced at $r_1 = 3$~fm. Center -- 4 tubes (v). Right -- 10 tubes (vi).}
\end{figure}

\begin{figure}[!htb]
\begin{center}
\includegraphics[width=60.0mm]{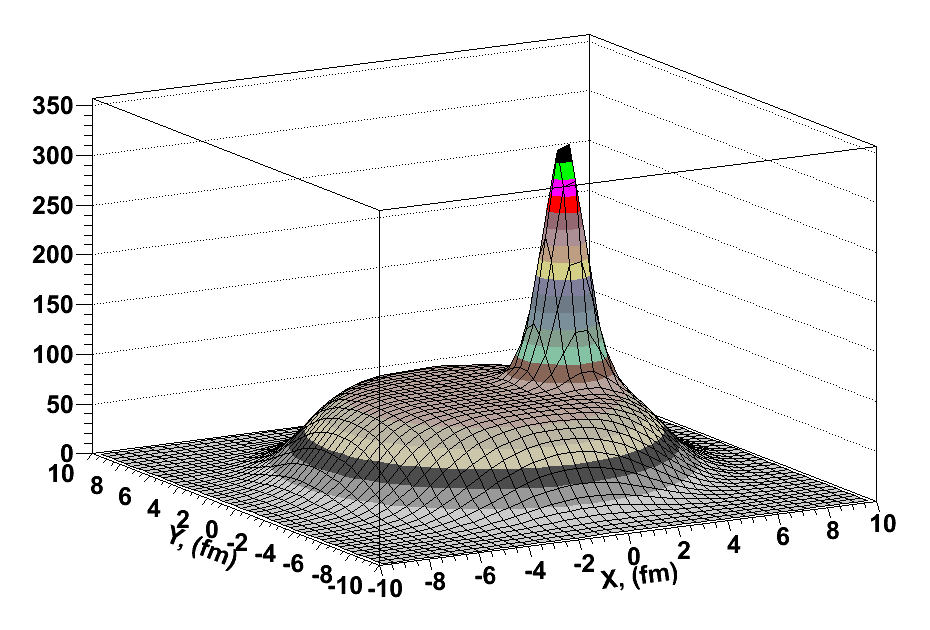}
\end{center}
\caption{3D plot of the initial energy density profile with one tube-like fluctuation (vii) placed at $r_i = 0.5R$ on the Woods-Saxon background (3).}
\end{figure}

The configurations described above serve as the initial conditions for hydrodynamic evolution of the superdense system. The quark-gluon plasma and hadron gas are supposed
to be in complete local equilibrium above the chemical freeze-out temperature $T_{ch}$ with EoS described below. With the given IC, the evolution of thermally and chemically equilibrated matter is described with the help of the ideal hydrodynamics approximation. The latter is based on (2+1)D numerical hydrodynamic code, described in \cite{Sinyukov2}. For this stage of evolution we use the lattice QCD-inspired equation of state of quark-gluon phase \cite{laine} together with corrections for small but non-zero baryon chemical potentials \cite{Sinyukov2}, matched with chemically equilibrated hadron-resonance gas via crossover-type transition.
The hadron-resonance gas consists of all ($N=329$) well-established hadron states made of u,d,s-quarks, including $\sigma$-meson, $f_0$(600). 
The chemical freeze-out hypersurface at $T = 165$~MeV, which corresponds to the end of chemically and thermally equilibrated evolution and start of the hydrokinetic stage of A+A collision process, is presented in Fig. 3 for the initial configuration~(iii).
 
     At the temperatures below $T_{ch}$ the system loses chemical and thermal equilibrium and gradually decays. In the hydrokinetic approach, HKM, the dynamical decoupling is described
by the particle escape probabilities from the inhomogeneous hydrodynamically expanding system (with resonance decays taken into account) in a way consistent with the kinetic equations in the relaxation-time approximation
for emission functions \cite{Akkelin, Sinyukov2}. The HKM gives the possibility to calculate the momentum spectra formation during continuous process of the particle liberation.  Within HKM one can consider the single event (= fixed IC) basing on numerical calculations of the analytical formulas for spectra using the temperatures and particle concentrations from numerical hydrodynamic solution. An extensive description of the approach can be found in \cite{Akkelin,Sinyukov2}. In the employed version of HKM there is no Monte Carlo cascade algorithm for the hadron stage, that typically brings the new ensemble of events  with energy-momentum conservation fulfilled only in average for the fixed initial conditions and fixed hydrodynamic evolution (in this aspect the results of some studies do not correspond, strictly speaking, to the true event-by-event analysis). We just calculate the analytical structure describing gradual decay of expanding fluid into the particles in the single event (related to the fixed IC) in the  sense that solutions of the Boltzmann equation describe continuously the {\it mean} distribution functions at given IC in correspondence with  the energy-momentum conservation law. 

In this paper the HKM is applied to the systems with the bumpy ICs which are described above.   
\begin{figure}[!htp]
\begin{center}
   \includegraphics[width=110mm]{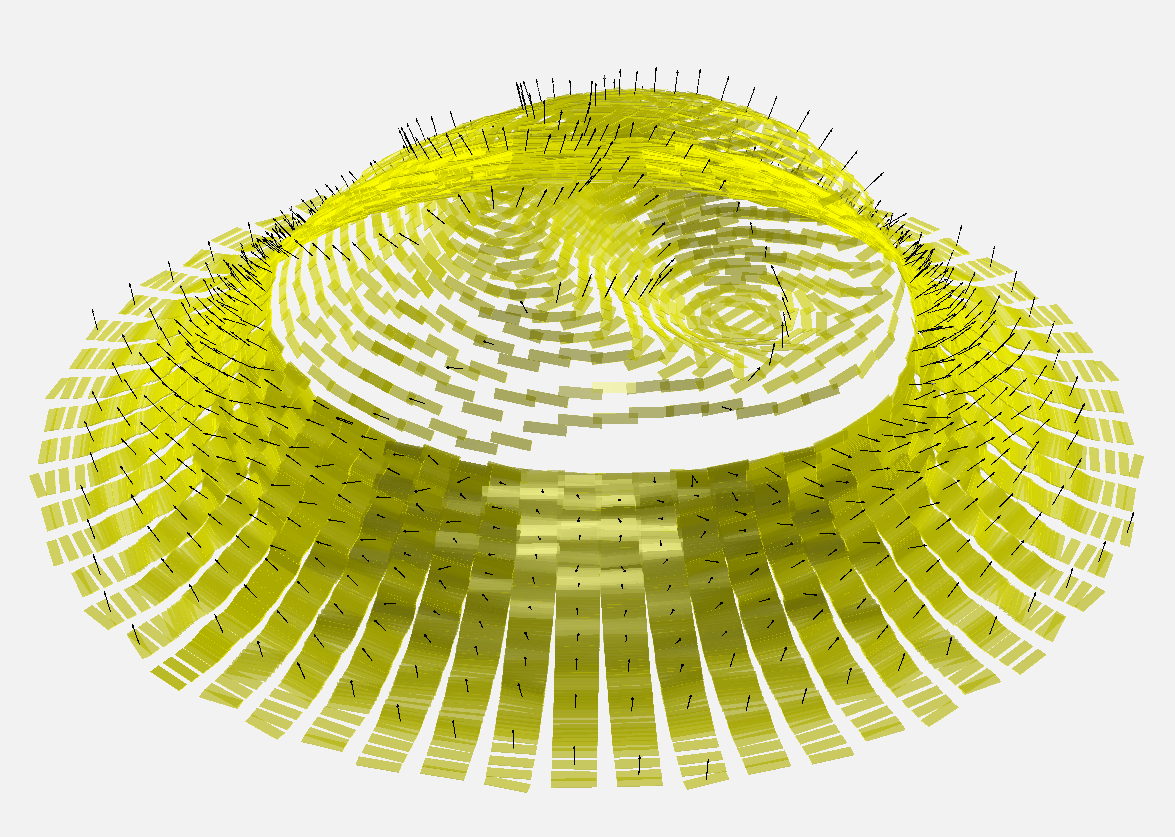} 
\caption{The chemical freeze-out hypersurface $\tau(x, y)$ at $T = 165$~MeV for 1 tube (iii) placed at $r_1 = 3$~fm from the center with the vector field of velocities.}
\end{center}
\end{figure}

\section{Results and discussion}

In this Section we present the results obtained within HKM model with the tube-like IC. 
It is obvious, that the angular dependence of transverse pion spectra in the cases (i) and (ii), described in the previous Section, is flat, and the effective temperature of the $p_T$-spectrum (its inverse slope) is higher for configuration (ii). A non-trivial angular dependence appears for the other ICs, where at least one tube is shifted from the center and so brings an azimuthal asymmetry into the system. In Fig. 3, which fully corresponds to the actual calculations, one can see  that the fluctuation in the initial distribution leads to the appearance of concavity on freeze-out hypersurface in the azimuthal direction corresponding to the initial high-energy fluctuation. It means, in particular, that the freeze-out temperature is reached earlier in this domain. 

\textit{3.1. Integrated Pion Spectra.} The analysis of the spectra at various $p_T$ shows their angular dependence to be different. In this subsection we demonstrate the results for integrated over $p_T$ spectra, aiming to see a possibility for the ridge formation in such a tubular picture.  In Fig.~4  we present $|p_T|$-integrated pion spectra $dN/d\phi$  for different initial configurations. The black line corresponds to the integrated spectra at $p_T  > 0.9$ GeV and the red one corresponds to the momentum values $p_T > 2.5$ GeV 
\footnote{It is demonstrated in different hydrodynamic models, see e.g. \cite{KarpSinWern}, that the hydrodynamics works in $p_T$-region until 3 GeV for the central events; here we consider very central collisions with $b\approx 0$. Note that we analyze the contribution of only hydro-component to the soft ridge formation.}.  
One can see that behavior of the curves in the case of the Woods-Saxon background distribution with the single very peripheral tube (Fig.~4, $2^{nd}$ row, right) is similar to the results obtained in \cite{Hama1} for similar configuration; in particular, both local minima coincide at $\phi$=0, but in contrast to \cite{Hama1}, for $p_T> 2.5$~GeV the minimum is not absolute but local.  This probably  points out that Woods-Saxon background is not similar to the one used in \cite{Hama1}, for the Nexus-inspired IC. For the Gaussian-like background or not so peripheral single-tube disposition (Fig.~4, top left) the angular behavior of the spectrum is different from the above result. Nevertheless, the azimuthal positions of the maximal values of the spectra are correlated for high momentum ``trigger'' component and ``associated'' soft one. In the $3^{rd}$ row of Fig.~4 for three initial tubes there are no positive correlations in a front of the initial tube at $\phi=0$. Nevertheless, there are such correlations at the left figure at the points $-\pi$ and $\pi$ and at the right figure at $-1$ and 1 rad. The angular positions of the maximal values are synchronized for 4 and, partially, for 10 tubes (Fig.~4, bottom). 

\begin{figure}[!htp]
\begin{center}
 \begin{tabular}{cc}
 \includegraphics[width=60mm]{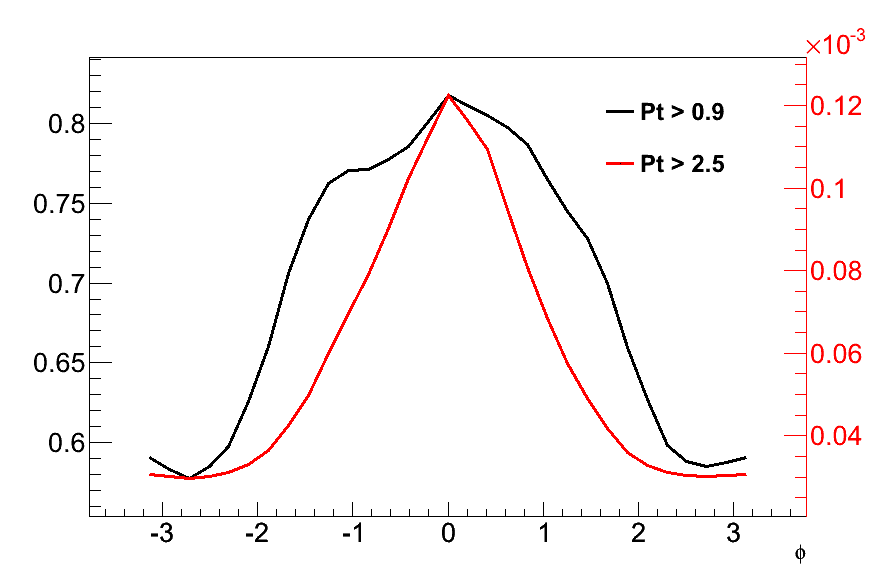}&
 \includegraphics[width=60mm]{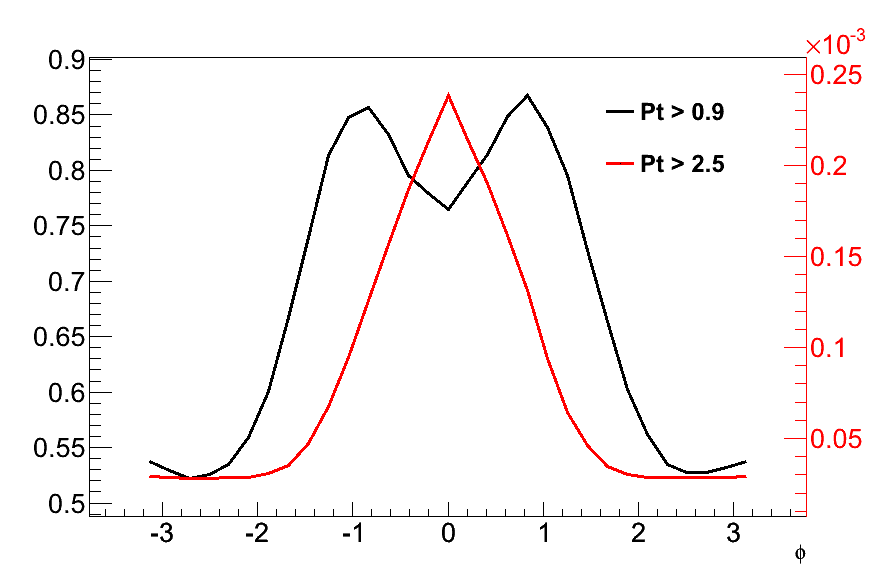}\\
 \includegraphics[width=60mm]{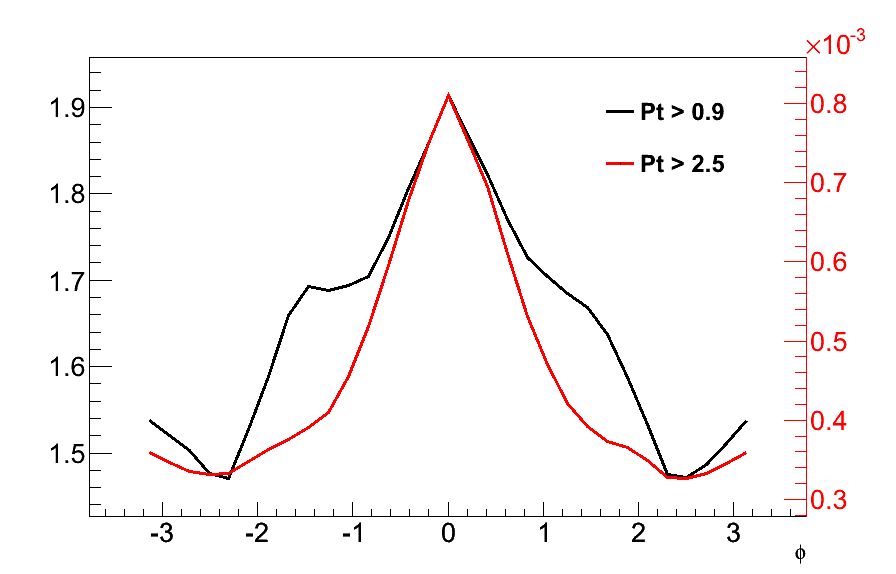}&
 \includegraphics[width=60mm]{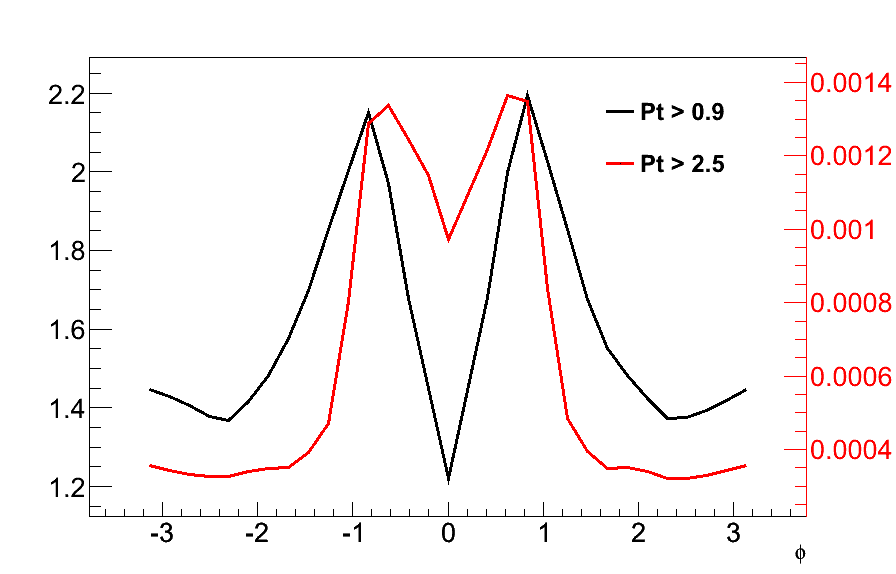}\\
 \includegraphics[width=60mm]{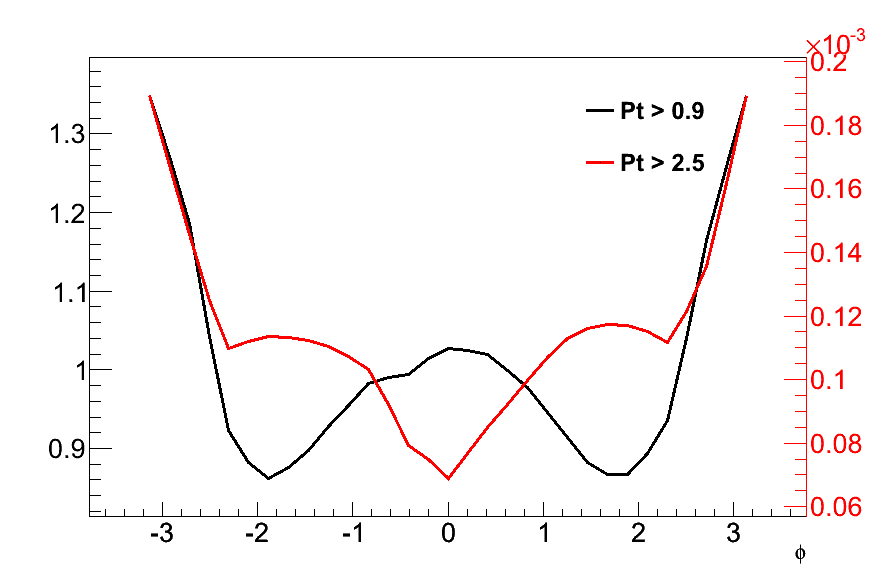} &
 \includegraphics[width=60mm]{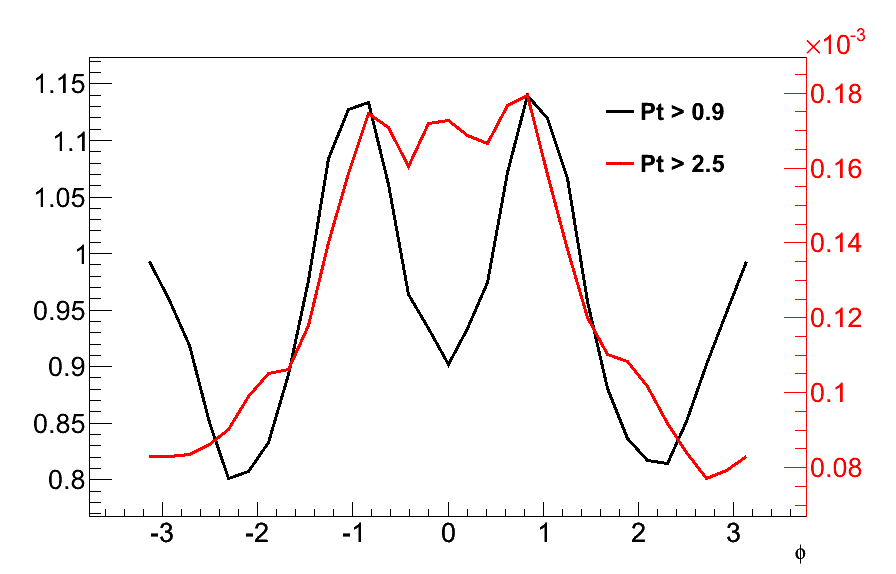}\\
 \includegraphics[width=60mm]{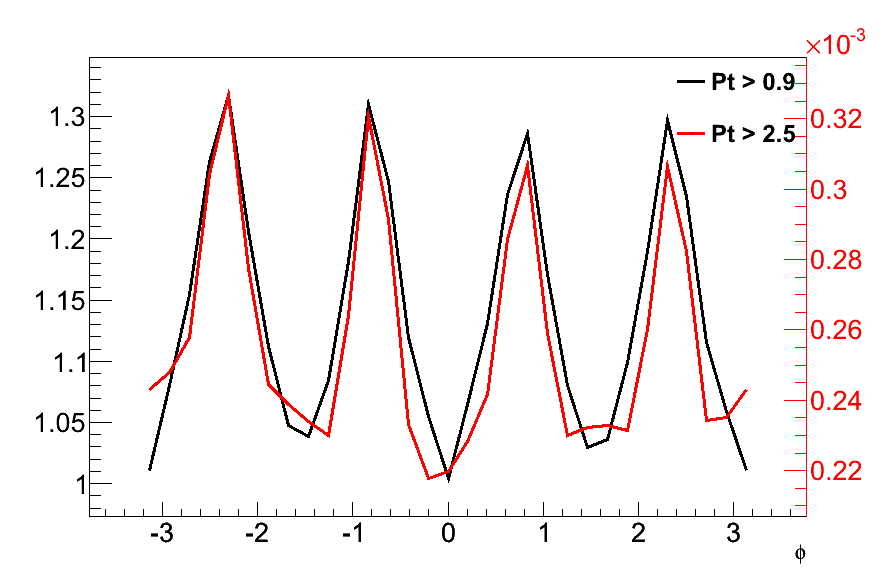}&
 \includegraphics[width=60mm]{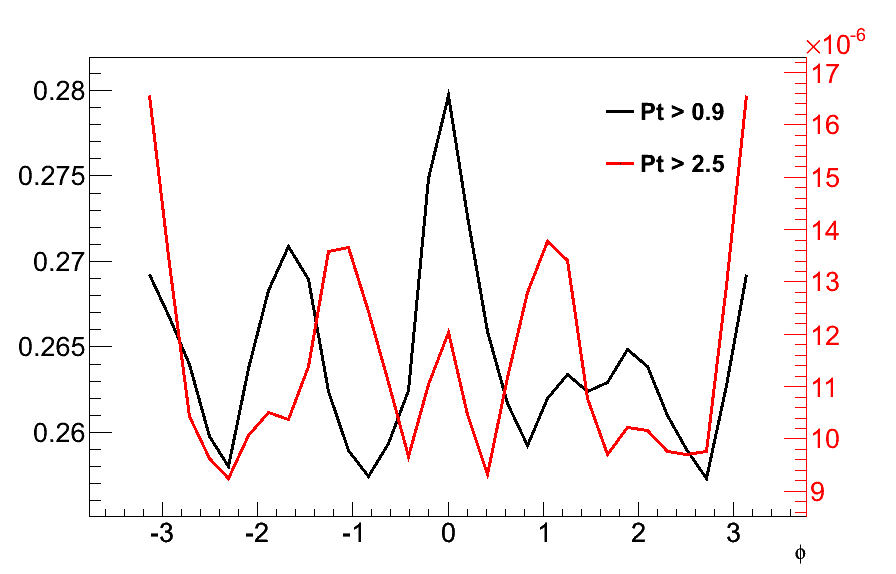}\\
\end{tabular}
\caption{The integrated spectra with the tube-like ICs. The top row demonstrates the case of 1 displaced tube with the Gaussian background (iii):  left -- at $r_1 =0.55R$, right -- at $r_1 =1.04R$. The $2^{nd}$ row corresponds to the case of 1 displaced tube with the Woods-Saxon background (vii):  left -- at $r_i =0.5 R$, right -- at $r_i =1.1 R$. The $3^{rd}$ row is related to the IC with 3 tubes and the Gaussian background (iv): left -- $\textbf{r}_i =(0, 0)$,$(-1, 3.6)$,$(-1,-3.6)$~fm, right -- $\textbf{r}_i = (0, 5.6)$,$(-1, 3.6)$,$(-1, -3.6)$~fm. The $4^{th}$ row is related to the Gaussian background: left -- 4 tubes (v), right -- 10 tubes (vi).}
\end{center}
\end{figure}

\begin{figure}[!htp]
\begin{center}
 \begin{tabular}{cc}
 \includegraphics[width=57mm]{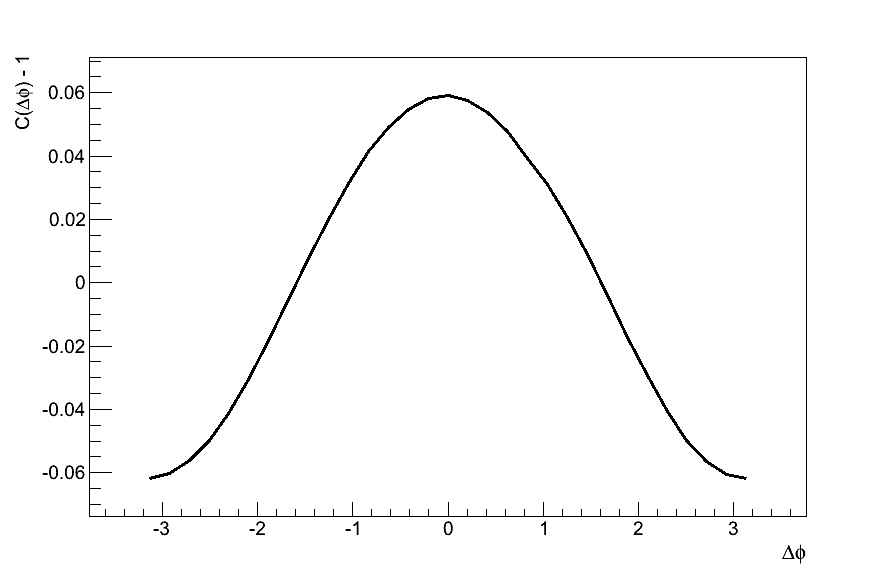}&
\includegraphics[width=57mm]{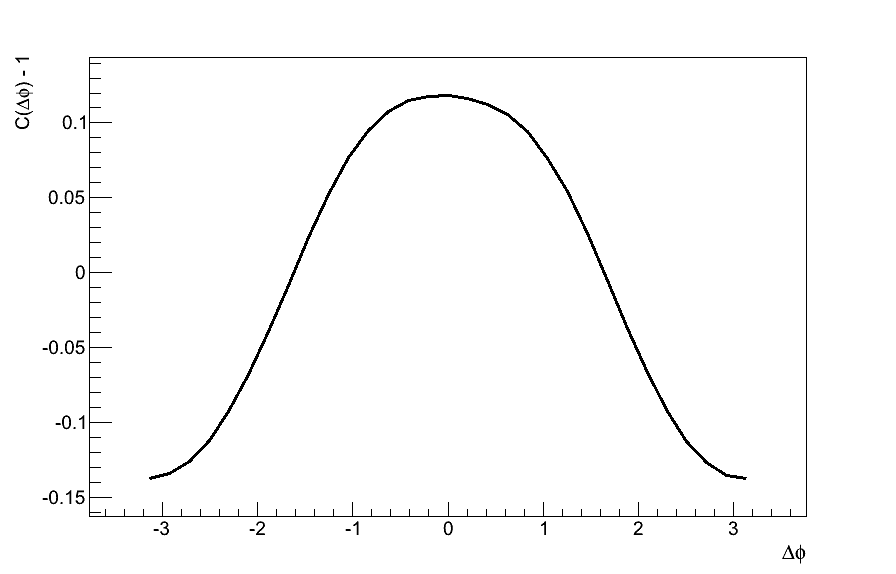}\\
 \includegraphics[width=57mm]{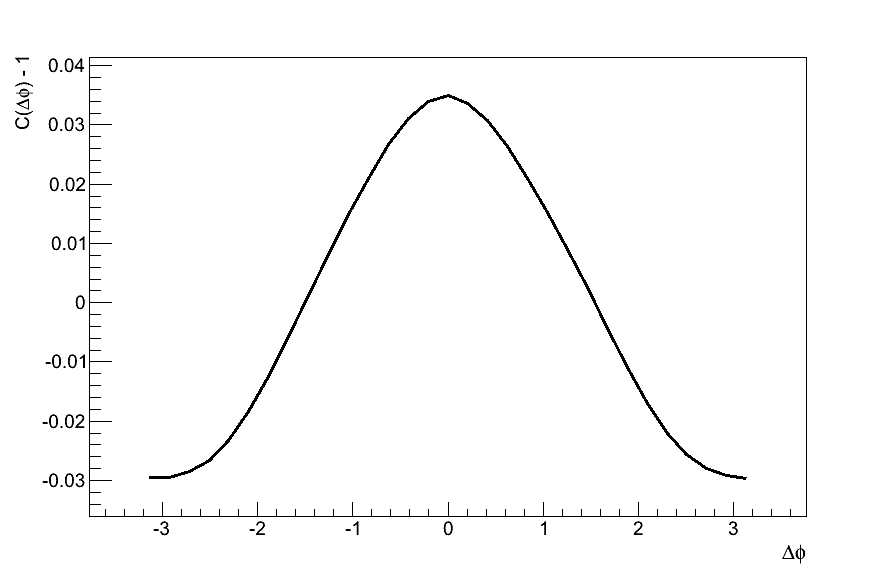}&
 \includegraphics[width=57mm]{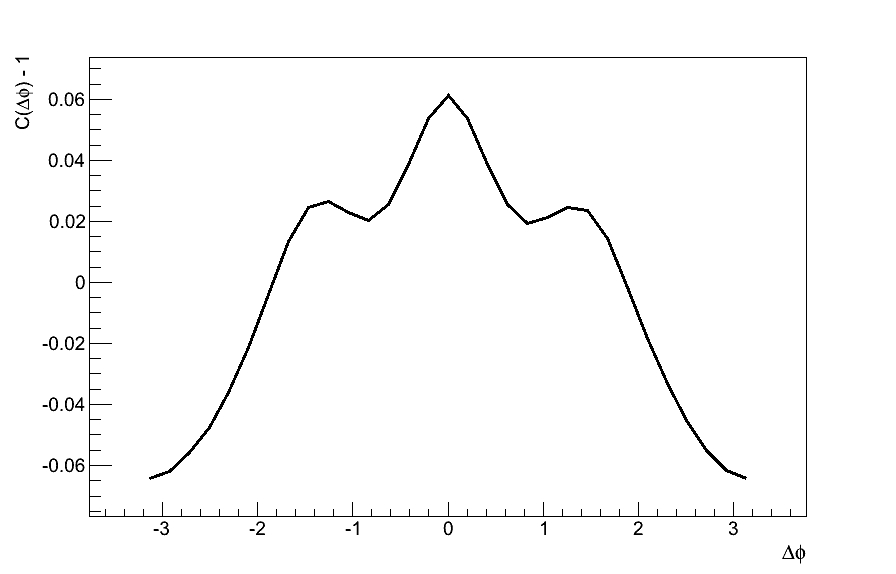}\\
 \includegraphics[width=57mm]{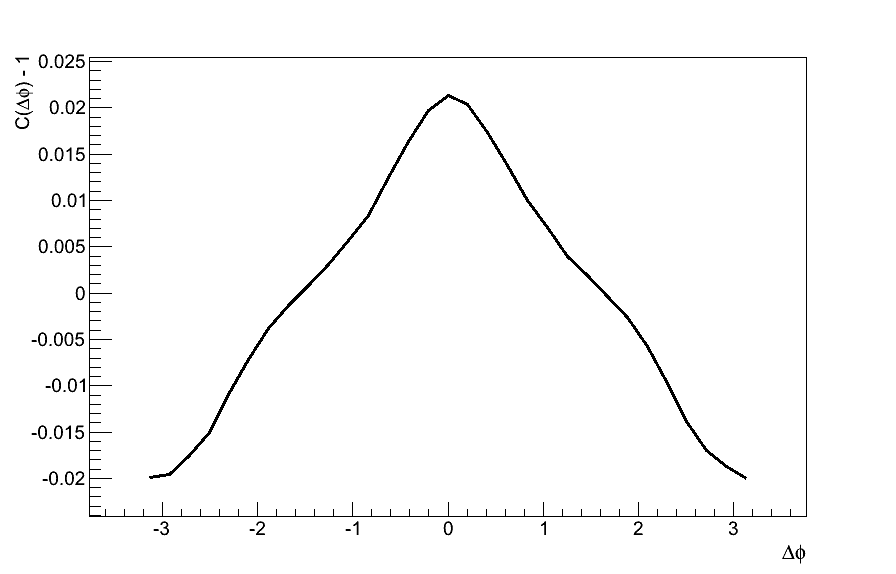}&
 \includegraphics[width=57mm]{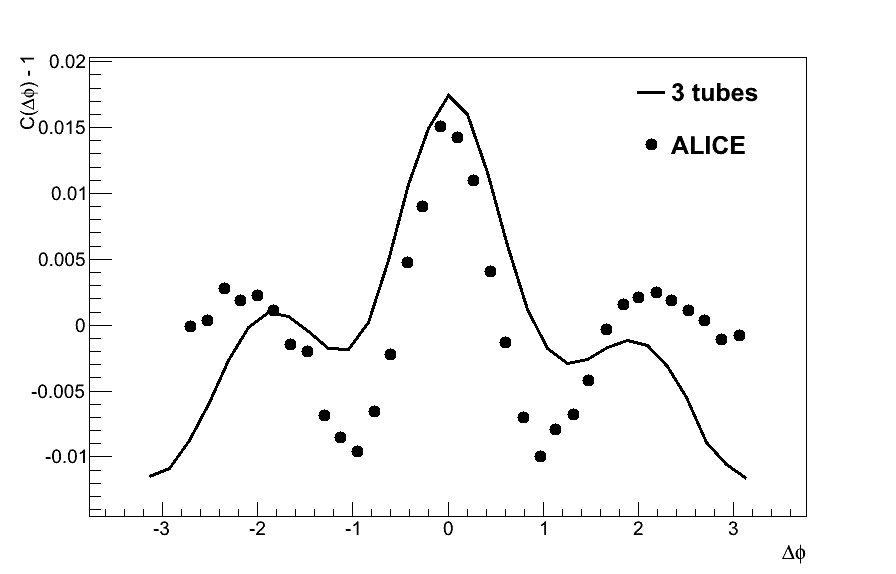}\\
 \includegraphics[width=57mm]{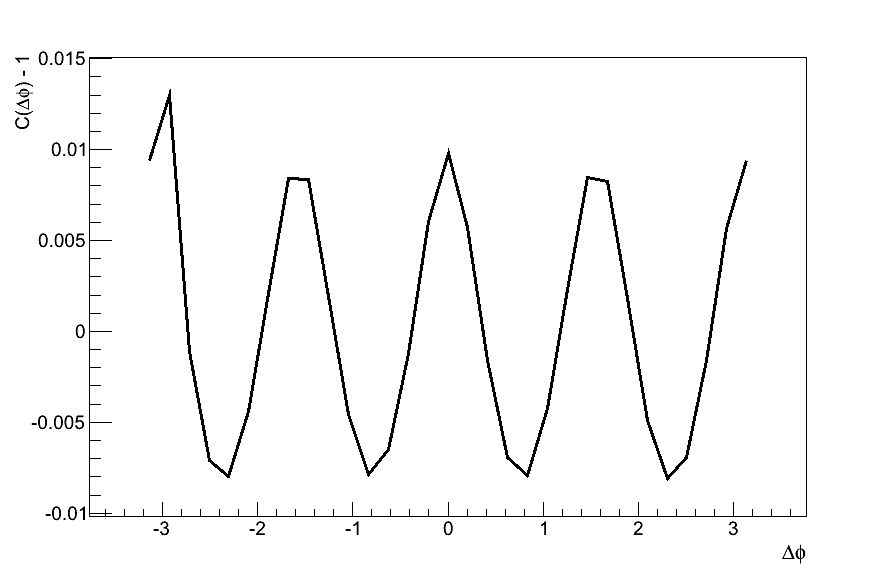}&
 \includegraphics[width=57mm]{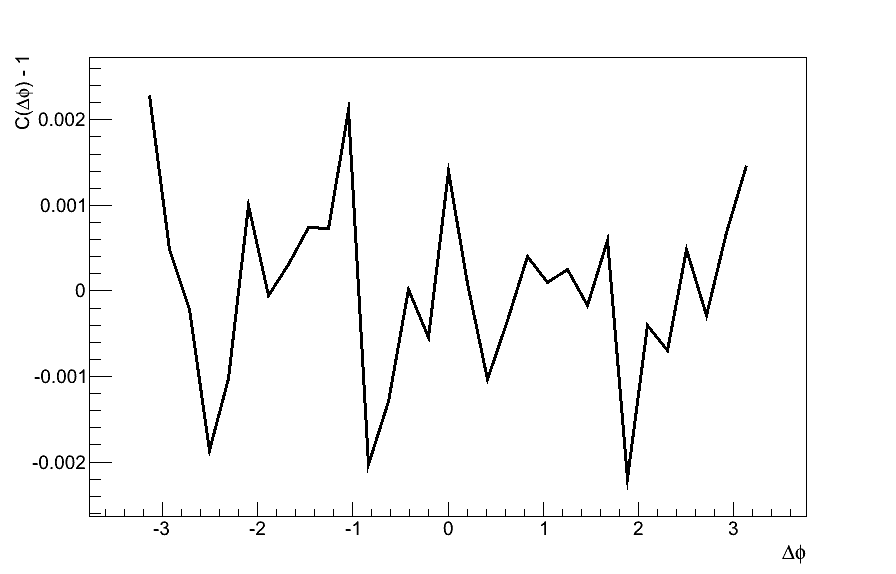}\\
\end{tabular}
\caption{The correlation function $C(\Delta\phi)$ for various ICs. The top row demonstrates the case of 1 displaced tube with the Gaussian background (iii):  left -- at $r_1 =0.55R$, right -- at $r_1 =1.04R$. The $2^{nd}$ row corresponds to the case of 1 displaced tube with the Woods-Saxon background (vii):  left -- at $r_i =0.5 R$, right -- at $r_i =1.1 R$. The $3^{rd}$ row describes 3 tubes with the Gaussian background (iv): left -- $\textbf{r}_i = (0, 0)$,$(-1, 3.6)$,$(-1,-3.6)$~fm, right -- $\textbf{r}_i = (0, 5.6)$,$(-1, 3.6)$,$(-1, -3.6)$~fm. The $4^{th}$ row is related to the Gaussian background: left -- 4 tubes (v), right -- 10 tubes (vi). The ALICE LHC data are taken from  \cite{ALICE}.}
\end{center}
\end{figure}

The results demonstrate that if one particle with relatively large $p_T$ is triggered, then, most probably, its azimuthal direction will correspond to the one of the peaks in the distributions $dN/d\phi$. Then, as it follows from the $dN/d\phi$ distributions for $p_T> 0.9$~GeV and $p_T> 2.5$~GeV, the probability to find the second ``associated'' particle with the same or smaller transverse momentum will be maximal in a narrow range $\Delta\phi$ near this peak. Such effects are typically expressed through the ratio $C(\Delta\phi)$ of the dihadron distribution in $\Delta \phi$ for the {\it same} events to the one extracted from the {\it mixed} events:

\begin{equation}
C(\Delta\phi)-1=\left(\frac{dN^{mixed}}{d\Delta\phi}\right)^{-1}\left(\frac{dN^{same}}{d\Delta\phi}-\frac{dN^{mixed}}{d\Delta\phi}\right),
\end{equation}
where in our model for each identical (in the sense of variable geometry analysis) initial tube-like configuration we have 

\begin{eqnarray}\label{edd4}
&&\frac{dN^{same}}{d\Delta\phi}=\int\limits_{-\pi}^{\pi} f\left(\phi+\frac{\Delta\phi}{2}\right)\cdot g\left(\phi-\frac{\Delta\phi}{2}\right) d\phi, \nonumber \\
&&\frac{dN^{mixed}}{d\Delta\phi}=\frac{1}{2\pi}\int\limits_{-\pi}^{\pi}\int\limits_{-\pi}^{\pi} d\phi_1 d\phi_2  f\left(\phi_1+\frac{\Delta\phi}{2}\right)\cdot g\left(\phi_2-\frac{\Delta\phi}{2}\right)=\frac{1}{2\pi}N_{as}N_{tr}.
\end{eqnarray}
Here $f(\phi) =\int\limits_{0.9}^{3.0} s(p,\phi) dp$ and $g(\phi) =\int\limits_{2.5}^{3.0} s(p,\phi) dp$  are $dN/d\phi$ distributions for $p_T> 0.9$~GeV and $p_T> 2.5$~GeV respectively, $s(p,\phi)$ are the pion spectra, and $N_{tr}=\int\limits_{-\pi}^{\pi} g(\phi)d\phi$ and $N_{as}=\int\limits_{-\pi}^{\pi} f(\phi)d\phi$ are the  normalization constants for the ``trigger'' and ``associated'' components respectively.

The correlation functions $C(\Delta\phi)$ are presented in Fig.~5 for the different initial configurations. One can see that for many of considered initial configurations (but not for all of them!)  there are narrow azimuthal near-side correlations which are typical for the ridge. The inclusive correlations between the triggered particle and the particle corresponding to the other peaks will be relatively weak, since these peaks will change their angular location from event to event with respect to the ``triggered'', or near-side, peak, and this will wash out the two-particle correlations outside the near-side peak. It is not unlikely, however, that the doubly-peaked away-side structure, which is seen in Fig.~5 ($2^{nd}$ row right and $3^{rd}$ right) can be supported by the multi-peak configurations similar to those presented in  Fig.~5 (bottom), and so the doubly-peaked away-side structure may survive, if the weights of such triangular initial configurations are fairly big. One can also see that if the tube is not in the IC peripheral region, just the near-side ridge structure appears, and the away-side structure is washed out.  Thus, not only the number of the tubes is important, but also their positions. The comparison of the ridge and doubly-peaked away-side structure with the data of ALICE LHC \cite{ALICE} supports such a hypothesis. This is the possible mechanism of formation of the so-called soft ridge structure~\footnote{
Note, that the dihadron correlations are sensitive not only to the tube configurations, but also to the tube energy, as it is analyzed for one-tube case in \cite{Hama2,Hama1}.}. 
The comparison between the results for the Gaussian background (Fig.~5, $1^{st}$ row, right) and the
Woods-Saxon one (Fig.~5, $2^{nd}$ row, right) demonstrates sensitivity of ridge formation to the background 
initial profile: in the case of one tube the ridges prefer Woods-Saxon type of the profile.

As for the formation of the hard ridges with triggered particle momenta $p_T> 5-6$~GeV, it is very likely that they are the result of superposition of the bulk spectrum structure, caused by the tubular bumpy IC, and the jet spectrum one, conditioned by the jet formation mechanism accounting for the interaction with the bulk matter. Even a 10\%   coincidence between the azimuth jet direction and the angular position of one of the peaks in soft spectra, similar to those in Fig.~4, can be enough to form the observed hard ridge \cite{Werner}.    

\textit{3.2. The Flow Harmonics and  $p_T$ Dependence of Their Magnitudes.} The anisotropy of the initial tubular conditions that leads to non-trivial ridge structure, results also in a non-trivial transverse momentum spectrum angular structure, even in the most central collisions. This structure can be  investigated quantitatively using a discrete Fourier decomposition. Then the azimuthal momentum distribution of the emitted particles is commonly expressed as
\begin{equation}\label{edd6}
\frac{dN}{p_T dp_T d\phi dy} = \frac{1}{2\pi} \frac{dN}{p_T dp_T dy} \left(1+ \sum\limits_{n=1}^{\infty}  2 v_n(p_T) cos(n(\phi-\Psi_n(p_T)))\right),
\end{equation}
where $v_n$ is the magnitude of the $n^{th}$ order harmonic term, relative
to the angle of the initial-state spatial plane of symmetry $\Psi_n$.
In Fig.~6 we show our results for the $p_T$ dependence of $v_n$ for different initial configurations. 
Here, since our purpose is to show clearly the effects of bumpy IC, we plot the results obtained for $v_n$ up to $n=6$. 
One can note that $v_1$ coefficients take negative and positive values,
and so the integral $\int_0^{3.5 \mathrm{GeV}}dp_T p_T^2 v_1(dN/p_Tdp_T)$ is close to
zero. In fact, this integral is less than $10\%$ of the differences
between its values in the regions where $v_1>0$ (large $p_T$)
and $v_1<0$ (small $p_T$). This smallness of the integral (for
different configurations it is $0.004-0.029$~GeV) reflects the
transverse momentum (it is zero initially) conservation. Since
we consider $v_n$ coefficients for the pion subsystem only, one
cannot expect exact zero value for considered integral.

Note that for configurations (ii)--(vii) for odd number
of the initial tubes the odd harmonics are dominating, and
the analogous tendency takes place for the even harmonics
at even number of the tubes. However, to make general
conclusion, more configurations have to be considered.

The experimentally valuable results have to be based on some model of the tube-like initial conditions and the correspondingly built event generator for IC, and also include an event by event evolution procedure. It will define the weight of the single events, such as those presented in Fig. 6, and finally give the resulting event-averaged $v_n$-coefficients and the quantitative structure of the ridge. We plan to realize such a quantitative experimental analysis in subsequent studies.  

\begin{figure}[!htp]
\begin{center}
 \begin{tabular}{cc}
  \includegraphics[width=55mm]{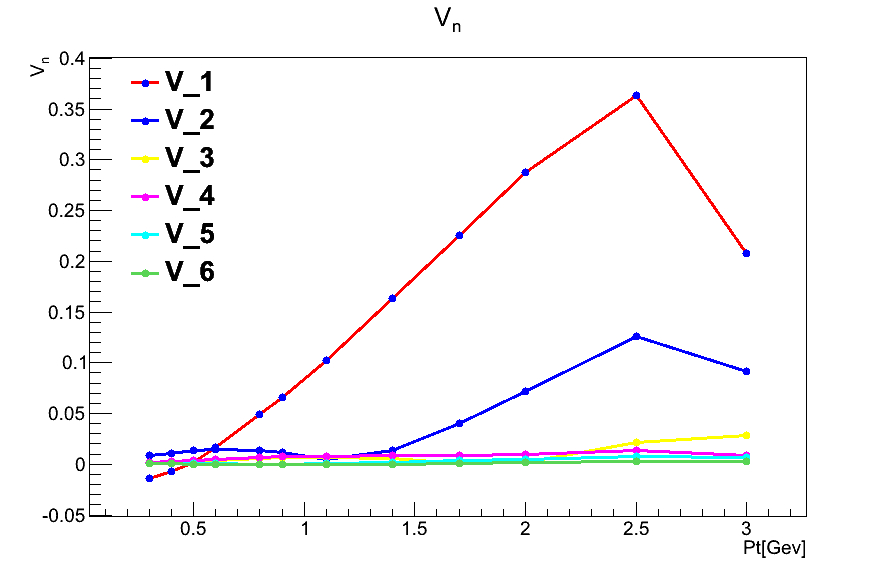}&
  \includegraphics[width=55mm]{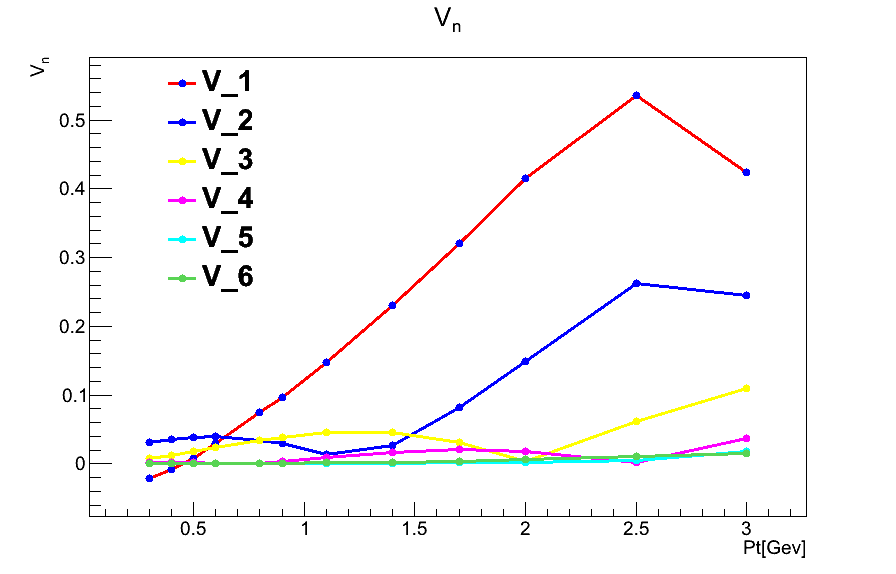}\\
  \includegraphics[width=55mm]{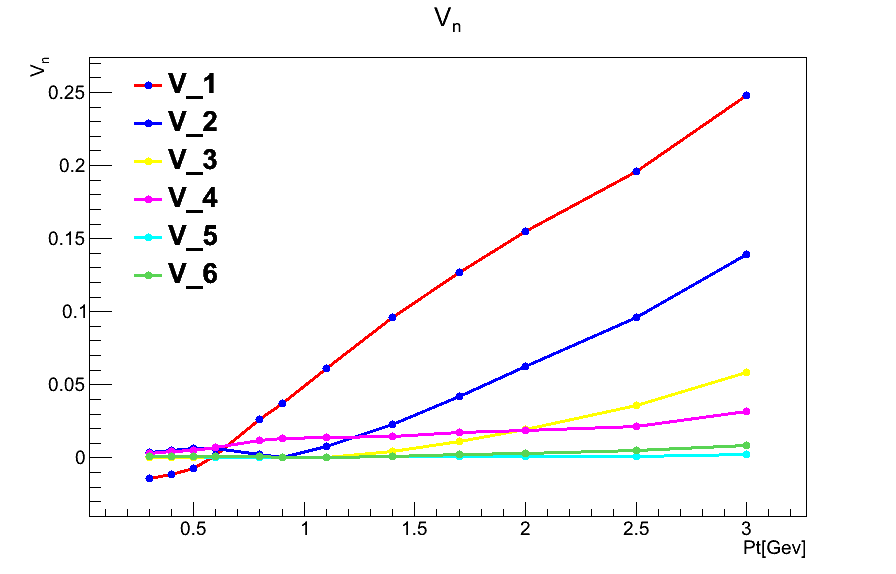} &
  \includegraphics[width=55mm]{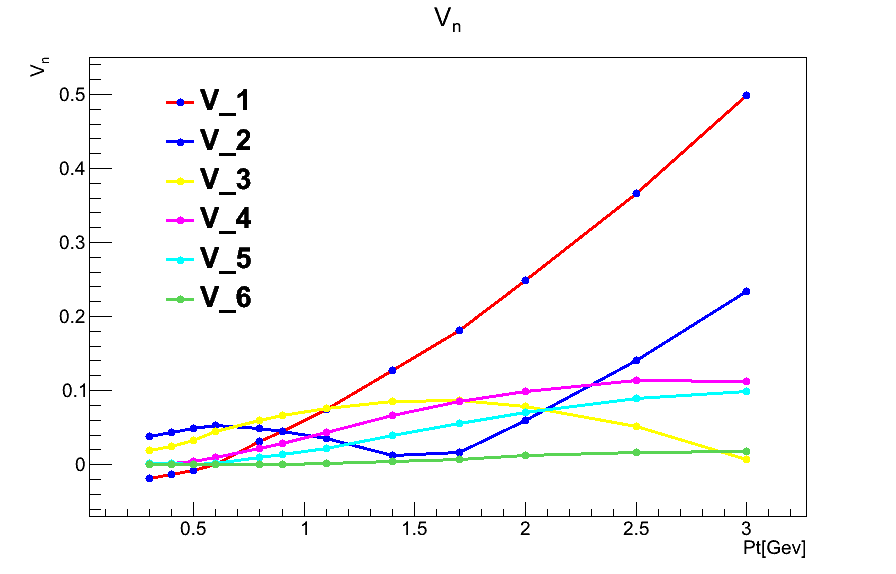}\\
  \includegraphics[width=55mm]{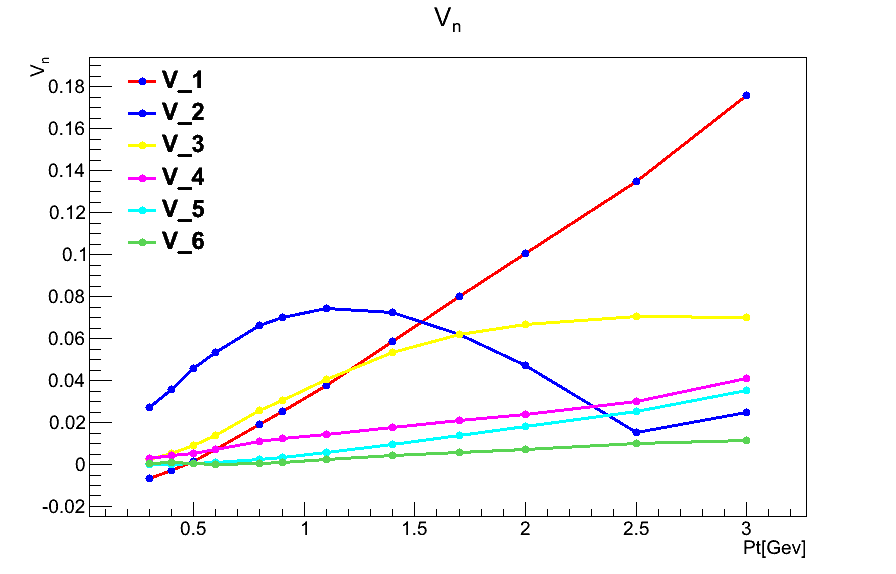}&
  \includegraphics[width=55mm]{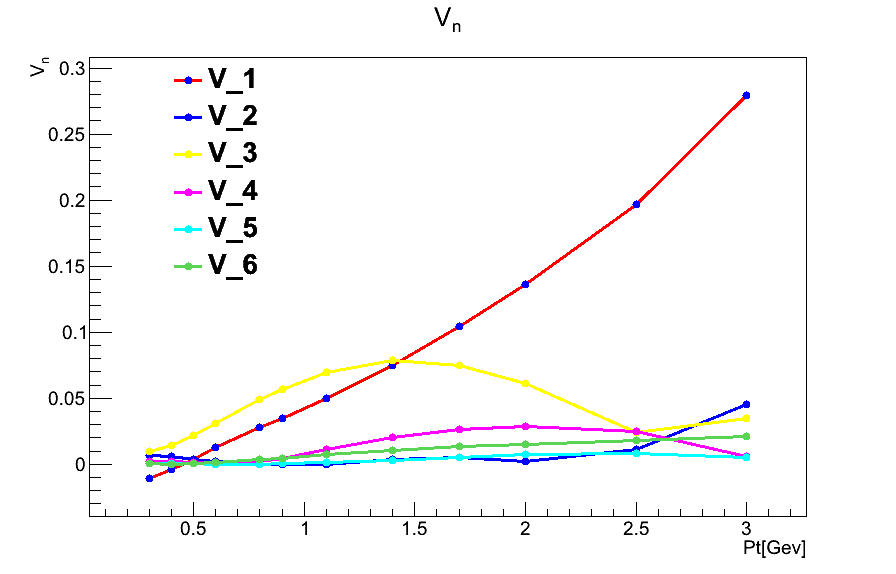}\\
  \includegraphics[width=55mm]{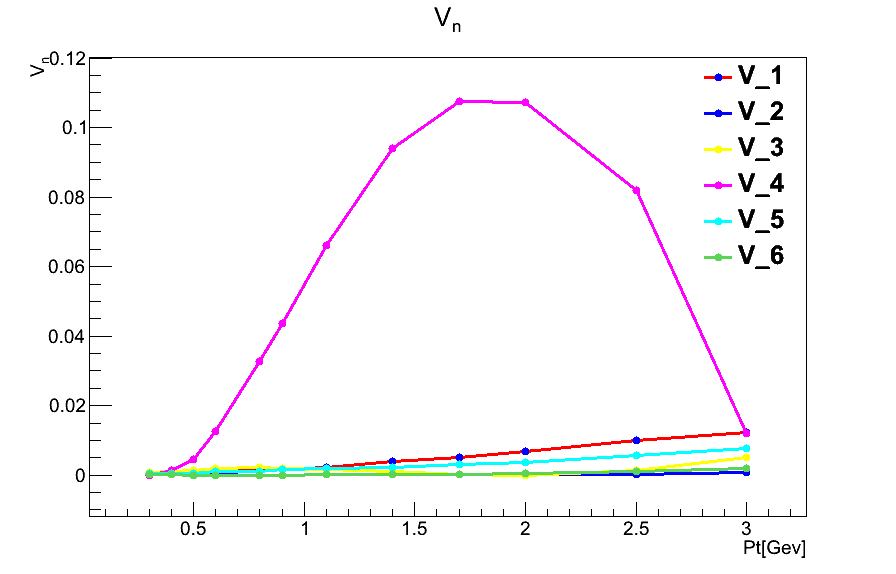} &
  \includegraphics[width=55mm]{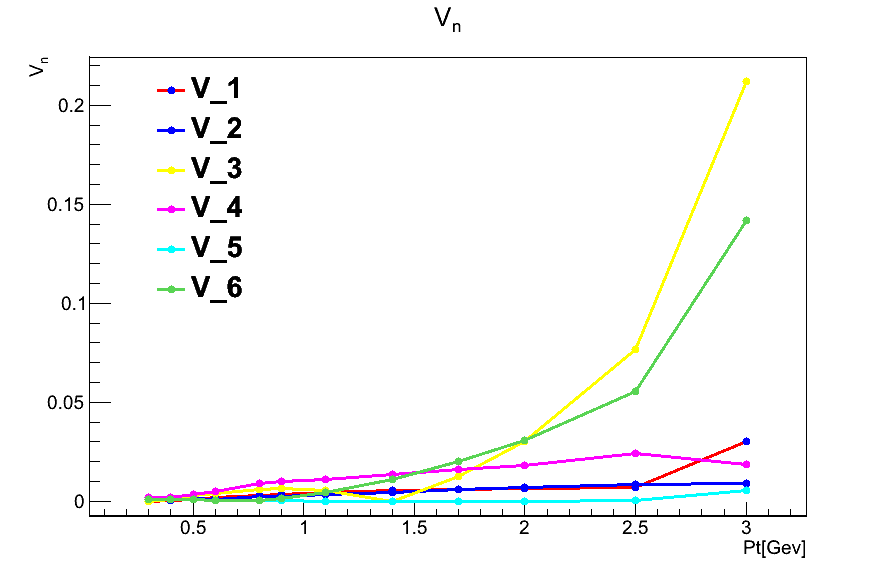}\\ 
  \end{tabular}
\caption{The $v_n$-coefficients for different tube-like ICs. The top row demonstrates the case of 1 displaced tube with the Gaussian background (iii):  left -- at $r_1 =0.55R$, right -- at $r_1 =1.04R$. The $2^{nd}$ row represents 1 displaced tube on the top of the Woods-Saxon distribution (vii):  left -- at $r_i =0.5 R$, right -- at $r_i =1.1 R$. The $3^{rd}$ row corresponds to 3 tubes on the top of the Gaussian distribution (iv): left -- $\textbf{r}_i = (0, 5.6)$, $(-1, 3.6)$, $(-1, -3.6)$~fm, right -- $\textbf{r}_i = (0, 0)$,$(-1, 3.6)$,$(-1,-3.6)$~fm. The $4^{th}$ row is related to: left  -- 4 tubes  with the Gaussian background (v), right -- 10 tubes  with the Gaussian background (vi).} 
\end{center}
\end{figure}

\section{Conclusions}
The tube-like fluctuating structures in the initial energy density distribution are considered with the aim to study the influence of their presence on the pion spectra, flow harmonics and ridge formation. These very dense color-field flux tubes are formed at the very initial stage of the nucleus-nucleus collision and lead to the long-range longitudinal correlations in pseudorapidity. It is found that the presence of the corresponding bumpy structures in transverse direction in IC strongly affects the hydrodynamic evolution and leads to emergence of conspicuous structures in  azimuthal distributions of the pion transverse momentum spectra. The hydrokinetic evolution  for different initial configurations with different numbers of tubes is calculated. As the result, one can see that most configurations can bring the ridge structure accompanied by the specific set of  $v_n(p_T)$-coefficients. Not only the triangular structures of the initial conditions are responsible for  the soft ridge formation, but also the odd number of the initial tubes can support this job. It means that  the hydrodynamic mechanism of the near-side ``soft ridges'' formation is sufficiently plausible. Also one can note that the doubly-peaked away-side structure appears when there is one outer/peripheral tube in the initial conditions.  To constrain event by event fluctuating IC in A+A collisions with subsequent hydrodynamic expansion within the viscous HKM and provide the detailed quantitative analysis of the ridge structure a further systematic analysis is planned.

\section{Acknowledgments}
Yu.~M.~Sinyukov thanks Y. Hama for the fruitful discussions and FAPESP (Contract 2013/10395-2) for financial support.
The research was carried out within the scope of the European Ultra Relativistic Energies Agreement (EUREA) of the European Research Group GDRE: Heavy ions at ultrarelativistic energies and is supported by the State Fund for Fundamental Researches of Ukraine, Agreement F33/24-2013 and the National Academy of Sciences of Ukraine, Agreement  F4-2013. Iu.~A.~Karpenko acknowledges the financial support by the ExtreMe Matter Institute EMMI and Hessian LOEWE initiative.

\end{document}